\documentclass[prl,aps,showpacs,twocolumn,superscriptaddress,amsmath,amssymb,floatfix]{revtex4}

\usepackage{graphicx}
\newcommand{\be}{\begin{equation}}
\newcommand{\ee}{\end{equation}}
\newcommand{\xx}{{\mathbf x}}

\newcommand{\kk}{{\mathbf k}}

\newcommand{\eq}[1]{(\ref{eq:#1})}
\newcommand{\eqname}[1]{\label{eq:#1}}
\newcommand{\mume}{\mu \rm m}

\newcommand{\ahd}{\hat{a}^\dagger}
\newcommand{\ch}{\hat{c}}
\newcommand{\chd}{\hat{c}^\dagger}
\newcommand{\bh}{\hat{b}}
\newcommand{\bhd}{\hat{b}^\dagger}
\newcommand{\ket}[1]{|#1\rangle}

\newcommand{\braopket}[3]{\langle#1|#2|#3\rangle}

\begin{document}
\title{Fermionized photons in an array of driven dissipative nonlinear cavities}
\author{I. Carusotto}
\affiliation{BEC-CNR-INFM and Dipartimento di Fisica, Universit\`a di Trento,
I-38050 Povo, Italy}
\affiliation{Institute for Quantum Electronics, ETH Z\"urich, 8093 Z\"urich, Switzerland}
\author{D. Gerace}
\affiliation{Institute for Quantum Electronics, ETH Z\"urich, 8093 Z\"urich, Switzerland}
\affiliation{CNISM-UdR Pavia and Dipartimento di Fisica ``A. Volta'', Universit\`a di Pavia, 27100 Pavia, Italy}
\author{H. E. T\"ureci}
\affiliation{Institute for Quantum Electronics, ETH Z\"urich, 8093 Z\"urich, Switzerland}
\author{S. De Liberato}
\affiliation{Laboratoire Mat\'eriaux et Ph\'enom\`enes Quantiques, Universit\'e Paris Diderot-Paris 7 and CNRS, UMR 7162, 
75205 Paris Cedex 13, France}
\affiliation{Laboratoire Pierre Aigrain, \'Ecole Normale Sup\'erieure, 24 rue Lhomond, 75005 Paris, France}
\author{C. Ciuti}
\affiliation{Laboratoire Mat\'eriaux et Ph\'enom\`enes Quantiques, Universit\'e Paris Diderot-Paris 7 and CNRS, UMR 7162,
75205 Paris Cedex 13, France}
\author{A. Imamo\v{g}lu}
\affiliation{Institute for Quantum Electronics, ETH Z\"urich, 8093 Z\"urich, Switzerland}

\begin{abstract}
We theoretically investigate the optical response of a one-dimensional
 array of strongly nonlinear optical microcavities. 
 When the optical nonlinearity is much larger than both
 losses and inter-cavity tunnel coupling, the non-equilibrium steady
 state of the system is reminiscent of a strongly correlated
 Tonks-Girardeau gas of impenetrable bosons. Signatures of strong
 correlations are identified in the absorption spectrum of the system,
 as well as in the intensity correlations of the emitted light. Possible
 experimental implementations in state-of-the-art solid-state devices
 are discussed.
\end{abstract}

\maketitle

Strong correlations in quantum many body systems give rise to a number of striking phenomena and states of matter, ranging from superfluidity of liquid Helium and superconductivity of metals to fractional quantum Hall effect in two-dimensional electron gases. Most of the work in condensed matter physics so far focused on systems close to thermodynamic equilibrium. It is known however that even richer range of behaviors and phases is exhibited by systems where the steady state arises from a dynamical balance between drive and dissipation rather than from a thermodynamic equilibrium condition~\cite{non_eq_phase_trans, non_eq}. The study of non-equilibrium phenomena has been historically confined to classical models, such as driven diffusive lattice gases~\cite{drivendiffusive} and models of hydrodynamic stability \cite{drazin}. Another well-studied, prototypical example of a non-equilibrium phenomenon is lasing~\cite{lasing}. In contrast, non-equilibrium phases of {\it quantum} many-body systems remained largely unexplored. Early work in this direction has addressed dissipative magnetic chains~\cite{magnetic}, driven electron gases~\cite{electron}, and nonlinear optical systems~\cite{polar_superfl_exp,polar_superfl,noneq_josephson}.

Several papers have recently investigated the possibility of observing strongly correlated quantum phases in gases of interacting photons~\cite{review}, including Mott insulator to superfluid transition in cavity arrays~\cite{polariton_BH}, interacting spin models~\cite{spin}, the Tonks-Girardeau gas in a waveguide geometry~\cite{polariton_TG}, and quantum Hall states~\cite{QH}. The focus of almost all this prior work has been the realization of well-known and thoroughly studied equilibrium many-body systems in photonic media. 

In this Letter we report the theoretical investigation of a genuine non-equilibrium state of strongly correlated photons. We study the spectroscopic signatures of strong interactions in an array of cavities mutually coupled by tunneling and driven by a coherent laser field. The properties of the non-equilibrium steady state are studied as a function of the pump frequency and intensity for a range of system parameters. Specific attention is devoted to the strongly nonlinear case, where non-trivial quantum correlations appear between photons indicating the onset of a Tonks-Girardeau gas of ``fermionized photons''. Most significantly, because the quantum correlations inside the many-cavity system directly transfer to the emitted radiation~\cite{WM}, a much wider range of observables is experimentally accessible than in analogous systems of ultracold atoms~\cite{TGatoms}.

\begin{figure}[htbp]
\begin{center}
\parbox[c]{0.35\columnwidth}{\includegraphics[width=0.35\columnwidth,angle=0,clip]{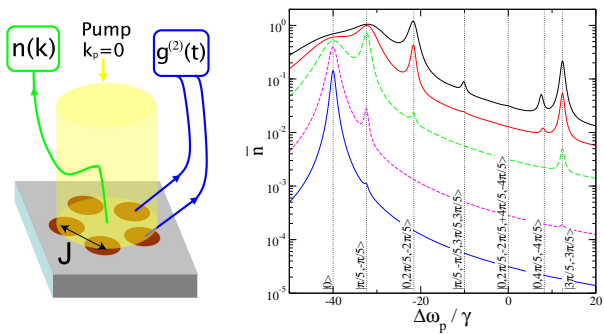}}
\parbox[c]{0.6\columnwidth}{\includegraphics[width=0.55\columnwidth,angle=0,clip]{figura_spectra_int.eps}}
\end{center}
\caption{Left panel: sketch of the system under consideration. 
%A necklace of $M$ (here $M = 5$) identical strongly nonlinear cavities is uniformly excited by a continuous wave laser of frequency $\omega_p$ at normal incidence $k_p=0$. 
%Each cavity mode has on-site repulsive energy $U > 0$, a loss rate $\gamma$ and is coupled to its two nearest neighbors via the tunneling  energy $J$. 
Right panel: absorption spectra of the mean number $\bar{n}$  of photons present in each cavity as a function of the frequency $\Delta\omega_p=\omega_p-\omega_0$ of the pump beam, taken at normal incidence $k_p=0$. System of $M=5$ cavities in the impenetrable boson limit $U/J=\infty$ with $J/\gamma=20$. 
Different curves correspond to increasing values of the pump amplitude $F_p/\gamma=0.1$, $0.3$, $1$, $2$, $3$. The vertical dotted lines indicate the spectral positions of the peaks predicted by the fermionization procedure with different number of photons. The kets $\ket{q_1\ldots q_N}$ indicate the pseudo-momenta $q_1 \ldots q_N$ of the occupied fermionic orbitals.}
\label{fig:spectra}
\end{figure}

The system we consider is sketched in the left panel of Fig.\ref{fig:spectra}. Assuming that the dynamics is restricted to a single photon mode per cavity, the Hamiltonian can be written in the following generalized Bose-Hubbard form:
\begin{multline}
\mathcal{H}=\sum_i \hbar \omega_0 \chd_i \ch_i + \hbar U\,\chd_i\chd_i\ch_i\ch_i -
\sum_{\langle i,j \rangle} \hbar J \, \chd_{i} \ch_{j} + \\ 
+\sum_i \left[ F_i(t)\,\chd_i + F_i^*(t)\,\ch_i \right] \, ,
\eqname{BH}
\end{multline}
where the $\ch_i$ ($\chd_i$) operators destroy (create) a photon in the $i$-th cavity located at $\xx_i$.
$\langle i,j \rangle$ indicates next-neighbor cavities, $U>0$ is the (repulsive) single-photon Kerr interaction within each cavity, $J$ is the inter-cavity hopping energy coming from the overlap of the nearest-neighbors cavity fields, and $\omega_0$ is the bare frequency of the isolated cavities. The term proportional to $F_i(t)$ accounts for the coherent drive by the pump laser: in the following we will restrict to the case of a monochromatic pump beam of frequency $\omega_p$, wavevector $\kk_p$, and amplitude $F_p$, i.e. $F_i(t)=F_p\,\exp[i(\kk_p \xx_i-\omega_p t) ]$.
Photons are assumed to be lost from the system at a spatially uniform rate $\gamma$. The master equation describing the time evolution of the density matrix $\rho(t)$ then has the standard Liouville form:
%\begin{equation}
${d_t\rho}=-i[\mathcal{H},\rho]+\frac{\gamma}{2}\sum_i (2 \ch_i \rho \chd_i-\chd_i \ch_i \rho - \rho \chd_i \ch_i)$.
%\eqname{master}
%\end{equation}

Let us start by discussing the eigenstates of the $N$ boson problem in the absence of pumping and dissipation.
In the linear regime $U=0$, photons occupy single-particle states of the hopping Hamiltonian, with an energy dispersion $\epsilon(k)=\omega_0-2J\,\cos(k)$. 
Wavevector is defined here as a dimensionless quantity: the first Brillouin zone (FBZ) then corresponds to the interval $k\in[-\pi,\pi]$.

In the opposite limit of impenetrable bosons $U/J=\infty$, a generic bosonic $N$-body wavefunction $\psi(i_1\ldots i_N)$ can be exactly mapped onto a fermionic wavefunction by the transformation~\cite{girardeau}:
\begin{equation}
\psi_F(i_1,\ldots,i_N)=\psi(i_1,\ldots,i_N)\,\epsilon(\sigma) \, .
\eqname{bose}
\end{equation}
Here, $i_1,\ldots,i_N$ are the positions of the $N$ particles, $\sigma$ is the permutation that sorts the spatial coordinates $i_1,\ldots,i_N$ into ascending order, and $\epsilon(\sigma)$ is the sign of the permutation $\sigma$. This sign guarantees that for any wavefunction $\psi$ symmetric under the exchange of any two particles, the corresponding $\psi_F$ is anti-symmetric as required by Fermi statistics.
%The impenetrability condition guarantees the vanishing of the wavefunction whenever any two particles coincide, i.e. $i_\alpha=i_\beta$.

As shown in~\cite{girardeau}, the eigenstates of the impenetrable boson problem are in a one-to-one correspondence with those of the non-interacting fermionic system, which are in turn simply classified by the occupation numbers of single-particle orbitals.
In the following, we shall use the shorthand notation $\ket{q_1\ldots q_N}$ to indicate the bosonic eigenstate corresponding to a Fermi wavefunction with one particle in each of the $q_1,\ldots,q_N$ orbitals. The (pseudo-)momenta $q_{\alpha=1\ldots N}$ are to be chosen within the FBZ, i.e. $q_\alpha\in[-\pi,\pi]$.
Both the energy and the {\em total} momentum of the bosonic state are equal to the ones of the corresponding fermionic one, say $E=\sum_\alpha \epsilon(q_\alpha)$ and $P=\sum_\alpha q_\alpha$ (as typical of a lattice, momentum is here defined only modulo $2\pi$).
On the other hand, the momentum distribution of the bosons is not preserved by the Bose-Fermi mapping. In particular, the {\em pseudo}-wavevectors $q_\alpha$ of the $\alpha=1\ldots N$ fermionic orbitals do not have a direct meaning in terms of physical observables of the bosonic system and in particular do not correspond to their physical momentum $k$~\cite{girardeau,olshanii}.

In what follows, we shall focus our attention on a necklace of $M$ cavities with periodic boundary conditions, i.e. $\psi(\ldots,i_\alpha=0, \ldots)=\psi(\ldots,i_\alpha=M,\ldots)$, $\forall \alpha$. A remarkable feature of the Bose-Fermi mapping is that the periodicity condition on the $N$-body bosonic wavefunction does not directly transfer to the single-particle orbitals of the fermionized wavefunction: depending on the number $N$ of particles in the system, the fermionic orbitals have to fulfill either periodic (if $N$ is odd) or anti-periodic (if $N$ is even) boundary conditions~\cite{LL,noiGutz}. 
This reflects on the quantization of the pseudo-wavevector $q=2\pi n/M$ (if $N$ is odd) or $q=2\pi (n+1/2)/M$ (if $N$ is even), where $n$ is an integer number. This peculiar quantization rule leads to the following explicit form of bosonic wavefunction of the lowest $N=2$ state $\ket{q,-q}$ with $q=\pi/M$:
\begin{equation}
\psi(i_1,i_2)=\frac{1}{\sqrt{2}\,M}\,\sin\left[\frac{\pi}{M}\,|i_1-i_2|\right] .
\eqname{N=2}
\end{equation} 

\begin{figure}[htbp]
\begin{center}
\includegraphics[width=0.9\columnwidth,angle=0,clip]{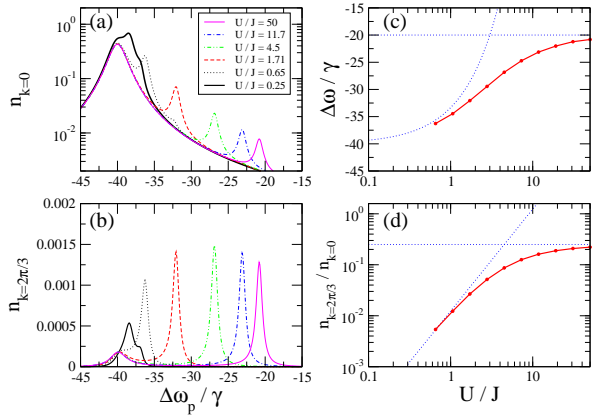}
\end{center}
\caption{
Panels (a,b) Spectra of the population in the $k=0$ and $k=2\pi/3$ bosonic modes as a function of pump frequency for a fixed pump amplitude $F_p/\gamma=0.5$ and different values of the nonlinear coupling $U/J$.
Panel (c,d): position of the peak and relative occupation $n(k=2\pi/3) / n(k=0)$ at the peak position as a function of $U/J$. Blue dashed lines: asymptotic values in the weak $U/J\ll 1$ and strong $U/J\gg 1$ interaction limits. System of $M=3$ cavities with $J/\gamma=20$. Pump at normal incidence $k_p=0$.
}
\label{fig:MF_SC}
\end{figure}

In contrast to equilibrium systems, pump and losses induce transitions between states with a different number of photons. This enables one to extract detailed information on the microscopic physics of the strongly interacting photon gas from the spectra of observable quantities. These can be numerically calculated by directly finding the stationary state of the master equation or by integrating it for long enough times via the Monte Carlo Wave Function technique~\cite{MCWF}. 
%The former method is generally more accurate, but quickly becomes untractable as the system size grows larger: the size of the superoperator in the density matrix space scales in fact as $(N_{max})^{4M}$ to be compared with the slower $(N_{max})^M$ scaling of the wavefunction.
Examples of absorption spectra are shown in Figs.\ref{fig:spectra} and \ref{fig:MF_SC}(a,b), where the mean number of photons per site $\bar{n}=\langle \chd_i \ch_i \rangle$ is plotted as a function of the pump frequency $\omega_p$ for a pump at normal incidence $k_p=0$ and a weak loss rate $\gamma \ll J,U$.

For very weak driving $|F_p|\ll \gamma$, the dynamics is mostly restricted to the vacuum state and the $N=1$, $\ket{q=0}$ state: the resonant driving of this transition is responsible for the main peak that is visible in all spectra at $\Delta\omega_p=-2J$.
For stronger driving amplitudes, higher-lying $N>1$ states start to appear. Photon number conservation implies that states containing $N$ photons in the many-cavity system is reached by repeated absorption of $N$ photons from the coherent drive. A generic many-body state $\ket{f}$ of energy $E_f$ will then appear in the spectra as a narrow resonance peak at frequency $\omega_p=E_f/N$.
In the impenetrable boson limit shown in Fig.\ref{fig:spectra}, the position of the peaks can be successfully compared to the analytical predictions of the Bose-Fermi mapping indicated by the vertical lines: each peak is associated to a set $q_{\alpha=1\ldots N}$ of pseudo-momenta that is compatible with momentum conservation $P=\sum_\alpha q_\alpha \equiv 0$.
Further confirmation of the peak assignments has been numerically obtained by looking at their power dependence: the intensity of a $N$ particle peak starts in fact as $|F_p|^{2N}$.  
On the other hand, even though the pseudo-wavevectors $q_\alpha$ completely identify a quantum state, they do not correspond to the physical momenta of the bosonic particles, so that no direct information can be inferred on the microscopic nature of the state from the momentum distribution of the corresponding emission.
%This crucial fact is easily visualized on the $N=2$ many-body wavefunction \eq{N=2}: the singularity at $i_1=i_2$ which makes the momentum distribution to have a long tail up to very large $k$'s.

In order to fully characterize the transition from the weakly to the strongly interacting regime, we have performed systematic numerical calculations for increasing values of the nonlinear coupling $U/J$ (Fig.\ref{fig:MF_SC}). These calculations have systematically and in full detail explored the $M=3$ case and we have checked that our findings extend in a straightforward way to larger systems with more sites. Spectra of the population $n_k=\langle \bhd_k \bh_k\rangle$ ($\bh_k=\sum_j e^{-ik j} \ch_j / \sqrt{M}$)  in the $k=0$ and $k=2\pi/3$ modes are plotted in the panels (a,b) for various values of $U/J$. Again, the main peak at $\Delta\omega_p=-2J$ corresponds to the resonance of the one particle $\ket{q=0}$ state. The additional peak that splits from it as $U/\gamma$ is increased corresponds to a two-particle final state. The $U/J$ dependence of its position and intensity (panels (c),(d)) provides insight into the microscopic nature of the two-particle state in the different regimes. 

In the strong interaction regime ($U/J \gg 1$), the two-particle state is well captured by the lowest $N=2$ fermionized state $\ket{q,-q}$ with $q=\pi/3$. This value of the pseudo-momentum imposed by the anti-periodic boundary condition is a clear signature of the fermionization effect and directly reflects into the asymptotic position of the peak at $\Delta\omega_p/J=-2\cos(\pi/3)=-1$. The relative intensity of the peak on the different modes $n_{k=\pm 2\pi/3}/n_{k=0}$ is also in perfect agreement with the analytical predictions $n_{k=\pm 2\pi/3}/n_{k=0}=1/4$. This latter value is obtained by discrete Fourier transform of the wave function \eq{N=2}. 

In the weakly interacting regime ($U/J\ll 1$), interactions can be treated within perturbation theory. To zeroth order in $U/J$, the lowest two particle state is a factorizable bosonic state with two particles in the $k=0$ mode and has an energy $-4J$. In this limit, it is visible only in the $k=0$ spectrum as a peak at $\Delta\omega_p=-2J$. 
At the next order, the state energy is blue-shifted by $2U/M$ and the wave function has the following analytical form:
\begin{equation}
\psi(i_1,i_2)\simeq \frac{1}{M}\left\{1-\frac{2U}{9\,J}\,\cos\left[\frac{2\pi}{3} ( i_1-i_2)\right] \right\}\, .
\eqname{N=2_dilute}
\end{equation}
The hole around $i_1=i_2$ that was complete in the fermionized wavefunction of the strong interaction limit \eq{N=2} is here much less pronounced and its depth scales as $U/J$. Correspondingly, the relative intensity of the peak at $\Delta\omega_p\simeq -2J+U/M$ on the $k=\pm 2\pi/3$ momentum components grows as $n_{k=\pm 2\pi/3}/n_{k=0}\simeq (U/9J)^2$.  Note how the interaction-induced blue-shift of the peak eventually saturates for large $U/J\gg 1$: the kinetic energy cost of creating a node at $i_1=i_2$ is well compensated by the suppressed interaction energy.

\begin{figure}[htbp]
\begin{center}
\includegraphics[width=0.8\columnwidth,angle=0,clip]{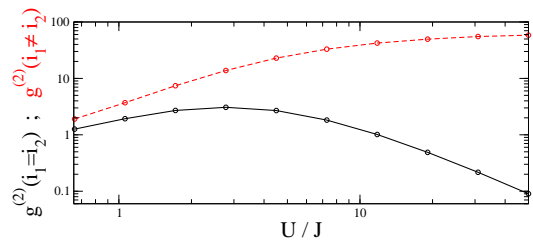}
\end{center}
\caption{Plot of the auto- (black, solid line) and cross-intensity correlations (red, dashed line) of the emission as a function of the interaction strength, $U/J$. Same system as in Fig.\ref{fig:MF_SC}.
}
\label{fig:g2}
\end{figure}

Further information on the microscopic nature of the many-body physics of the system can be obtained by inspecting the intensity correlations of the emitted light~\cite{WM} quantified by
%\begin{equation}
%g^{(2)}(i_1,i_2)=\frac{\langle \chd_{i_1}(t) \, \chd_{i_2}(t+\Delta t)\, \ch_{i_2}(t+\Delta t)\,\ch_{i_1}(t) \rangle}{\langle \chd_{i_1} \ch_{i_1}  \rangle\langle \chd_{i_2} \ch_{i_2}  \rangle}
$g^{(2)}(i_1,i_2)={\langle \chd_{i_1}\, \chd_{i_2}\, \ch_{i_2}\,\ch_{i_1} \rangle}/\bar{n}^2$
%{\langle \chd_{i_1} \ch_{i_1}  \rangle\langle \chd_{i_2} \ch_{i_2}  \rangle} \, .
%\end{equation}
Both the auto- ($i_1=i_2$) and the cross- ($i_1\neq i_2$) correlations are plotted in Fig.\ref{fig:g2} as a function of $U/J$ for a pump laser kept exactly on resonance with the ($U/J$-dependent) two-photon transition. 
In a very weakly interacting ($U/J\ll 1$) system, the two-particle peak overlaps the one-particle one, and the emitted light inherits the Poissonian nature of the pump laser.

For intermediate values of $U/J\approx 1$, the two-particle peak is already well separated from the single particle peak ($U/M \gg \gamma$). The resonant pump laser then selectively excites the two-particle state and a measurement of the intensity correlations of the emission provides detailed information on the quantum correlations between photons in this state.
The fact that the system preferentially contains $N=0,2$ particles rather than $1$ is responsible for the strong bunching. The almost flat shape of the wavefunction \eq{N=2_dilute} makes this bunching observable in both the auto- and the cross-correlations of the emission.
On the other hand, when interactions are very strong ($U/J\gg 1$) and the two-particle state has the fermionized form \eq{N=2}, the cross-correlation remains strongly bunched, but the auto-correlation becomes strongly anti-bunched as a consequence of the strong on-site interactions.

The basic building block of a possible device consists of a single cavity with a sufficiently strong on-site single-photon non-linearity. Although anti-bunched light generation has been shown for a number of emitter-cavity combinations such as atoms in high-finesse cavities~\cite{antibunching}, Cooper-pair boxes in super-conducting strip-line cavities~\cite{wallraff}, single quantum dots in micro-pillar~\cite{reithmaier} and photonic crystal defect cavities~\cite{antibunching_dots}, scalability to multi-cavity devices still remains a challenging task. An important requirement is that the difference between the resonance frequencies of the cavities have to be kept below the energy scales $U$ and $J$. To overcome this difficulty, laterally patterned microcavity systems containing quantum wells as the nonlinear medium appear to be the most promising candidates. State-of-the-art micropillar~\cite{pillars} and laterally patterned microcavity~\cite{patternedcavities} structures are expected to provide a sufficiently large polaritonic interaction~\cite{polariton_blockade} for this purpose. In both cases, the emission is strongly peaked in the vertical direction and the spectral position of the polariton mode is to a large degree determined by the photonic confinement, which in turn can be precisely engineered. Finally, significant tunneling between cavities separated by distances up to the $\mume$ range has been theoretically anticipated in~\cite{sarchi} for laterally patterned cavities, and also experimentally demonstrated in microdisk~\cite{microdisks} or photonic crystal~\cite{atlasov08} coupled cavity systems. This makes possible the independent collection of light emitted from each cavity, in addition to the selection in the far field~\cite{laguerreseparation}.

In conclusion, we have theoretically investigated the spectroscopic signatures of a non-equilibrium, strongly correlated gas of photons in a one-dimensional array of nonlinear cavities with strong photon-photon interactions and periodic boundary conditions. 
We have described the imprint of the transition from a weakly interacting system to a strongly interacting one, on experimentally accessible observables. We find that the absorption spectra show well-separated many-body resonances whose position provides an unambiguous signature of photonic fermionization; these can be used to spectrally isolate a single many-body state. We show how the microscopic structure of the non-equilibrium steady state of the system can be inferred from the intensity correlations of the emission. 
%We have finally demonstrated that the required parameter regime is accessible to state-of-the-art technology. 
Finally, we believe the present work demonstrates the importance of coupled nonlinear optical cavity systems in the theoretical and experimental investigation of quantum many-body systems out of equilibrium.

IC is grateful to J. Dalibard and Y. Castin for stimulating discussions at an early stage of the work. We acknowledge useful discussions with A. Badolato.
% AI acknowledges support from an ERC Advanced Investigator grant.

\end{document}